\documentclass[preprint,prb,tighten,floatfix,showpacs]{revtex4}
\usepackage{graphicx}

\def\<{\langle}
\def\>{\rangle}

\def\be{\begin{equation}}
\def\ee{\end{equation}}

\begin{document}
\preprint{cond-mat} \title{ Entanglement Entropy of Random Fractional Quantum Hall Systems }

\author{B. A. Friedman, G. C. Levine* and D. Luna}

\address{Department of Physics, Sam Houston State University, Huntsville TX 77341}

\address{*Department of Physics and Astronomy, Hofstra University,
Hempstead, NY 11549}

\date{\today}

\begin{abstract}       The  entanglement entropy  of  the   $\nu = 1/3$ and  $\nu = 5/2$ quantum Hall states in the     
presence of short range random disorder has been calculated by direct   diagonalization.  A microscopic model of electron-electron interaction is used, spin polarized electrons are confined to a single Landau level and interact with long range Coulomb interaction.  For very weak disorder,  the values of the  topological  entanglement entropy are roughly consistent with expected theoretical results.  
By considering a broader range of disorder strengths, the entanglement entropy was studied in an effort to detect quantum phase  transitions.  In particular, there is a signature of a transition as a function of the disorder strength for the  $\nu = 5/2$ state.  Prospects for using the density matrix renormalization group to compute the  entanglement entropy for larger system sizes are discussed.   \end{abstract}

\maketitle
\section{Introduction}
This paper is a numerical study, using direct diagonalization, of the entanglement  entropy of fractional quantum Hall systems in the presence of a delta correlated random potential.  The entanglement entropy, quite distinct from the thermodynamic entropy, is the Von Neumann entropy of the reduced density matrix of a subsystem and is a quantitative measure of the entanglement of the subsystem with the system.   Our interest in this subject is two-fold; firstly, it has been proposed that entanglement entropy can be used as a tool to characterize fractional quantum Hall states.  More precisely, Kitaev and Preskill \cite{kitaevpreskill} and Levin and Wen\cite{levinwen} have shown, for a topologically ordered state, that the entanglement entropy of a subsystem obeys an asymptotic relation 
 \begin{equation}
S \simeq \alpha L - \gamma + O(\frac{1}{L}) + \ldots 
\end{equation}
where L is the  linear ÒsizeÓ of the subsystem (the area law ) and  $\gamma$ is a universal     
quantity, the topological entanglement entropy, the natural logarithm of the quantum dimension.  For this scaling law to apply, the system must be very large and the subsystem must be large (compared to a cutoff, but the subsystem must be small compared to the system).  This is a rather  formidable numerical requirement, however, there has been some success numerically \cite{haque,zozulya,ent1,morris,ent2,zozulya2} using (1) to extract the topological entanglement entropy of quantum Hall states.  One may hope, that by adding weak randomness, there may be less system size dependence and hence it will be easier to obtain the topological entanglement entropy.  Of course, by adding randomness, momentum conservation is destroyed and one cannot treat as large systems  by direct diagonalization.   In any case, it is of interest to see if the topological entanglement entropy can be calculated in the presence of weak disorder and to see if the values obtained are consistent with previous numerical  estimates.

The second motivation to undertake this study, is to see whether the entanglement entropy can be used to detect transitions between phases of quantum Hall systems.  For example, experimentally, it is well known that fractional quantum Hall states are particularly sensitive to disorder.  Can this sensitivity be detected in the entanglement entropy?
The two questions discussed above will be studied for 2 filling factors $\nu = 1/3$ in the lowest Landau level, representative of Laughlin states, and the 5/2 th  state in the second Landau level.   Currently, there is good evidence both experimentally and numerically \cite{nayak} that the essential physics of the 5/2  state is given by the Moore-Read wave function and thus the 5/2  state is representative of the more exotic states with non abelian statistics.  

The paper is then organized as follows: in the second section, the model and the numerical method are briefly described and the results for the topological entanglement entropy for weak disorder are discussed.  In the third section, the entanglement entropy is calculated as a function of disorder strength for a wider range of disorder to  determine whether  transitions between phases of Hall systems can be detected.  In the fourth section, some preliminary results using the density matrix renormalization group to calculate the entanglement entropy are described.  The fifth section is a summary and gives conclusions.  
In the final section, a recent alternative method \cite{lauchli} to obtain the topological entanglement entropy on the torus is discussed.

\section{Extracting the Topological  Entanglement Entropy for Weak Disorder}

The numerical method we have used is direct diagonalization applied to square (aspect  ratio 1) clusters with periodic boundary conditions (the square torus geometry).  The Landau gauge is used for the vector potential.  Spin polarized electrons are confined to a single Landau level and interact with a pure Coulomb potential.  One can approach the limit of very large system sizes through clusters of any fixed aspect ratio and since we are  concerned with quantum liquid states, aspect ratio one has been chosen.  This numerical approach has previously been used to study the entanglement entropy without a disorder potential\cite{ent1,ent2}.  The random potential\cite{ultrav} U(r) is taken to  be delta correlated i.e.   $<U(r)U(r')> = U_0\delta(r-r')$ and the disorder strength will be given in terms of a parameter $U_R=\sqrt{3U_0/2}$.  Since momentum is not conserved, one is limited to smaller system sizes then for a  disorder free system.  In  particular, the largest system size treated for  $\nu = 1/3$  is 10 electrons in 30 orbitals with a state space of approximately  $30 X 10^6$ and 14 electrons in 28 orbitals for  $\nu = 5/2$ with a state space of  approximately $40 X 10^6$.  
(This is in contrast to the disorder free case,   $\nu = 1/3$  13 electrons in 39 orbitals , and   $\nu = 5/2$  18 electrons in 36  orbitals, are relatively straightforward to treat). 

To calculate the entanglement entropy, we take a subsystem  consisting of $l$ adjacent orbitals  (recall in the Landau gauge, these orbitals consist of strips oriented along, say the y-axis, of ÒwidthÓ of order the magnetic length).  The reduced density matrix is straightforward to compute from the ground state wave function.  It is then diagonalized  giving the  eigenvalues $\lambda_j$  from which the l-orbital entanglement entropy $S(l)$ ,  $S(l)=-\sum_j{\lambda_j \ln{\lambda_j}}$  is  obtained.  This procedure is done for every realization of the random potential, the results are then averaged to give $<S(l)>$ where $ < >$ denotes average over the random potential.  The position of the subsystem has been fixed, that is, for say $S(l=3)$ the subsystem always consists of the 1st , 2nd and 3rd orbitals.  For the smallest systems (6 electrons in 18 orbitals) we have averaged over 1000 realizations of the random potential, for the largest systems we have averaged over as few as 10 realizations.  This choice was dictated by the time consuming nature of the larger calculations.

In figure 1, we have plotted the entanglement entropy vs. square root of $l$ for 10 electrons in 30  orbitals.  The green circles are for no disorder ( an average is taken over the 3 ground states with $k_y=5,15,25$), while the blue and red circles are for disorder strength
$U_R=0.05$ averaged over 10 and  100 samples respectively.  The error bars are given by the root mean square values of $S(l)$ i.e. $\sigma=\frac{\sqrt{<S(l)^2>-<S(l)>^2}}{\sqrt{N_s-1}}$ with $N_s$ the number of samples.  From the relation (1) we expect linear behavior vs.  $\sqrt l$ for subsystems small compared to the system size; this behavior is seen in figure 1.  In particular, the linear regime is  ÒlargerÓ for the disordered case, indicating a smaller finite size effect for a given system size. 

This suggests a linear fit to the initial part of  the S(l) vs.  $\sqrt l $ curve to obtain the topological entanglement entropy as  the negative of the y-intercept.  
The results of the fit are plotted in figure 2 for 
$U_R=0.05$.  (the number of fitted  values of $l$ was chosen to give a local maximum in the value of $R^2$).  The topological entanglement entropy $\gamma$  , is found to be $1.10 \pm  .070 $ while a similar fit for $U_R=0.01$ gives $\gamma =1.13 \pm  .078$   both values in excellent agreement  with the value for the Laughlin 1/3 state of $2( \ln{\sqrt{3}}) \approx 1.10$,\cite{fend}
the factor of 2  coming from the 2 boundaries of the subsystem.  
However, as will be discussed below, the excellent agreement may be fortuitous in that for the small system sizes considered $\gamma$  tends to be overestimated at this filling.  

The dependence of the  topological entanglement entropy on system  size for filling 1/3 is shown in figure 3  for 
$U_R=0.01$.  In this figure $\gamma$   is plotted vs 1/N (N=number of orbitals).  Clearly, it would be desirable even with disorder, to be able to treat larger system sizes.  Another approach to obtain the topological entanglement entropy is, for  a given $<S(l)>$, to do a linear  extrapolation in 1/N yielding $<S^*(l)>$.  $<S^*(l)>$ is then plotted vs $\sqrt l$   , a linear least squares fit is performed  and the y intercept gives -$\gamma$.  A plot of $<S^*(l)>$  vs.   $\sqrt l$ is shown in figure 4, for  $\nu = 1/3$ using systems with 21 to 30 orbitals to get the extrapolations.  The negative of the y intercept is given by 1.30 with an error of 0.24 (the 0.24 due to  the deviation of the fit from a line.  The error in the extrapolations to get $<S^*(l)>$ was not taken into account.)  This method also gives a topological entanglement entropy consistent with the Laughlin 1/3 state.

Turning now to the    $\nu=5/2$ filling at  $U_R=0.01$, figure 5 shows $\gamma$   calculated from the initial linear part of $<S(l)>$ 
(i.e. figure 2) vs. 1/N.  For the largest system size, 14 electrons in 28 orbitals  $\gamma \approx 1.5 $ considerably less then the $\gamma$   expected from the  Moore-Read state (  $\gamma_{M.R.} \approx 2.08$  for 2  boundaries\cite{fend}).
If the 1/N dependence is fit by a line, one finds at $N=\infty$  a  $\gamma$  of 2.34 with an error of   0.08.  However, without knowing the answer, one does not know to extrapolate in figure 5 but not to extrapolate in figure 3, where the largest system sizes give ÒacceptableÓ answers without extrapolation.  In figure 6 $<S^*(l)>$ vs  $\sqrt l $ is plotted for   $\nu =5/2$ where $<S^*(l)>$ was obtained through extrapolation of system sizes 24,26,28.  The  y-intercept of the linear fit gives a $\gamma$  of 2.58.  Of course, this is no great success, however one obtains better agreement with the expected value if the  ratio $\frac{\gamma_{5/2}}{\gamma_{1/3}}$      is considered.  Using the $<S^*(l)>$ method ($<S^*(l)>$ obtained from the 3 largest system sizes)  $\frac{\gamma_{5/2}}{\gamma_{1/3}} \approx 1.97$    compared to  $\frac{\gamma_{M.R.}}{\gamma_{Laughlin}} \approx 1.89$.   

In any case, it appears the expected more  ÒgenericÓ behavior with weak randomness is unable to overcome the advantage of additional system  sizes available to disorder free calculations.  
That is, by using the S*(l) method, and 2 more system sizes (without disorder), reference \cite{ent1} was able to get agreement, with the expected theoretical results,   within the error bars, for both $\nu=1/3$ and $\nu=5/2$.  Without disorder, fitting a line to the initial part of the curve $S(l)$ vs. $\sqrt{l}$ is problematical for $\nu=1/3$, since, in addition to a linear increase, there is superimposed oscillation (see the FSS (finite size scaling result) of fig. 1 (a) of ref. \cite{ent1}).  We suggest that a similar oscillation, though less pronounced when disorder is present, causes difficulty in extrapolation of the data presented in figure 3.  To get the points in figure 3, for the smallest system size, 6 electrons in 18 states, the first 4 $l$-values were used to get the best linear fit (i.e. minimize $R^2$) while for 10 electrons in 30 states, the 6 initial $l$ values were used.  From the extrapolated values in figure 4 the oscillation causes an overestimate of $\gamma$ in both cases.  On the other hand, for $\nu=5/2$ there is less oscillation in $S^*(l)$, see figure 3a of reference \cite{ent1} and figure 6 of the present paper.  To obtain the points in figure 5, for 10 electrons, 7 $l$-values were used, while for 14 electrons, 8-$l$ values were used in the fit.  Due to less oscillation and a greater number of $l$-values used it is perhaps not surprising that extrapolation of $\gamma$, obtained from the initial part of $S(l)$, is more successful for $\nu=5/2$ then an extrapolation at $\nu=1/3$.     

\section{Entanglement Entropy as a Function of Disorder Strength}
In this section, the entanglement entropy is studied for a wider range of disorder strengths.  Entanglement entropy has been used previously  to investigate the phase diagram of quantum Hall systems  as a function of interaction potential \cite{zozulya2}, 1 dimensional quantum spin systems with disorder \cite{refael}, one particle entanglement entropy for Anderson transitions\cite{chak}   
and to study the phase diagram of the 1 dimensional extended Hubbard model \cite{mund} .  In figure 7 a,b $l$-entanglement  entropy $<S(l)>$ vs $U_R$ is plotted for filling 1/3.  In figures 7a $<S(2)>$ is  shown, this  figure being representative of small subsystems $l$.   In figures 7 b $<S(12)>$ is  graphed, this figure characteristic of larger subsystems.  Both graphs show a strong decrease in the entropy with disorder, with $S(12)$ exhibiting a slightly sharper decrease. Especially in the $S(12)$ graph, the entropy appears to decrease and then level out at a disorder strength of approximately $U_R = 0.25$. Although it is hard to make a definite conclusion, this is at least consistent with Wan et al. \cite{wan} that sees a vanishing of the mobility gap for $U_R > 0.25$.   


Let us now turn to filling 5/2.  The same  sequence of graphs is presented in figure 8 a,b.   Here there appears to be, especially for the large $l$ graph, fig 8b, a transition at a disorder strength $U_R \approx 0.04$.  A natural interpretation of these graphs is a quantum phase transition from the Moore-Read state for disorder strength $U_R \approx 0.04$.  Previous numerical studies \cite{ent1,li,thomale1} indicate that the ground state for pure Coulomb potential (no disorder) is topologically ÒequivalentÓ to the Moore-Read state.    We therefore suggest that the sharp drop off in figure 8a and particularly 8b as contrasted to the smoother curves in  7a and 7b is a transition due to the destruction of the Moore-Read state by disorder.  A possible picture of this transition is the destruction of p-wave superconductivity of composite fermions \cite{read} by disorder.  That such a transition should happen at rather weak disorder is physically appealing \cite{mac}.     

In an effort to characterize possible phase transitions with disorder, we have calculated the variance  $<S(l)^2>-<S(l)>^2$. In figures 9 and 10, the variance  for $l=12$ is plotted for  $\nu= 5/2$ and $\nu= 1/3$, respectively.  For $\nu= 5/2$, figure 9, the variance is nominal through the transition region (other then for the anomalous behavior of 10 electrons in 20 orbitals).  In contrast, for $\nu= 1/3$, figure 10, there is a general increase of the variance starting at $U_R=0.05$  and reaching a plateau at $U_R \approx 0.2-0.25$ which may indicate a transition in this range, consistent with figure 7, and consistent with reference \cite{wan}.

\section{Preliminary DMRG studies of the entanglement entropy}
A common ÒthemeÓ of the previous sections  is the benefit of finding a method to access larger system sizes.  A possible method to do this, for  quantum Hall systems, is to use the density matrix renormalization group (dmrg) \cite{white}.  One expects, the number of states kept in the dmrg blocks, needs to scale as the exponential of the entanglement entropy of the block, for an accurate calculation.  Since by the area law entropy scales as the  $\sqrt{s}$ where 
$s$ is the number of sites in the block, the number of states kept needs to scale as  $e^{c \sqrt{s}}$.  The bad news  is that this depends on the exponential of the $\sqrt{s}$   , however,the good news is that it does not depend on the exponential of $s$ as in direct diagonalization.  Hence, at least in principle, one should (if one can avoid being stuck in local minimum) be able to treat larger system  sizes for quantum Hall systems  by dmrg \cite{shibata1,shibata2,feiguin,withrow}.  
In particular, reference \cite{feiguin} was able to accurately calculate ground state energies for $\nu = 1/3$ for up to 20 electrons and up
 to 26 electrons for $\nu=5/2$  in the spherical geometry.  In the spherical geometry 14 electrons at $\nu=1/3$ and 20 electrons at $\nu=5/2$ are accessible to direct diagonalization.  However, the excitation gap, a more difficult numerical quantity at $\nu=5/2$ was only accurately calculable by dmrg for up to 22 electrons,  1 "non-aliased" system size larger then that accessible to direct diagonalization. 
In this section, dmrg will be used to calculate the entanglement entropy for quantum Hall systems without disorder.  We will be content, in this preliminary study, to use dmrg to study a large system size still accessible to direct diagonalization, that is, 12  electrons in 36 orbitals in the n=0 and n=1 Landau levels.  

In table I we display, the ground state energy vs. $m$, the number of states kept in the block; the first column is for the lowest Landau level, the second for the second Landau level.  (the Madelung energy, which can be calculated exactly, is not included).  
\begin{table} \caption{Comparison of Dmrg and Direct Diagonalization Energies}
  \begin{tabular}{ccc}
  $m$ & N=0 & N=1 \\ 
 200&-3.3675&-2.4109\\
 300&-3.3691&\\
 400&-3.3699&-2.4178\\
 600&-3.3717&-2.4203\\
 700&-3.3723&\\
 800& &-2.4219 \\
 Extrapolation &-3.3739$\pm .005$&-2.4252$\pm .003$\\
 Exact &-3.3734&-2.4254\\
  \end{tabular}
  \end{table}

One sees for the lowest Landau level a fairly accurate result can be obtained even without extrapolation, for the n=1 Landau level, extrapolation is more important (we extrapolate in $1/m$).  Let us now consider the calculation of $S(l)$ (recall $l$ is the number sites in the subsystem used when calculating the entanglement entropy) .  All $S(l)$Õs are computed at the end of the calculation when the left and right blocks have equal number of sites (in addition, there are two sites in the middle\cite{shibata2}).  This makes the calculations more complicated (i.e. clearly it is easier to get $S(l)$ when there are $l$ sites in the block)  but it is necessary to get reliable results.  Figure 11 is a plot of $S(l)$ vs.  $\sqrt{l}$ up to $l$=8 for differing number of states in the blocks for 1/3 filling.  One notices that even for the smallest block sizes, dmrg does a good job in computing $S(l)$.  This is consistent with the dmrg  calculations of the entanglement entropy done by  Shibata \cite{shibata3}.   Turning now to 1/3 filling in the second Landau  level ($\nu  = 7/3$), figure 12 is a plot of $S(l)$ vs.   $\sqrt{l}$ for this filling.  One again sees, that differing from the first Landau level, extrapolation is very important to obtain an accurate result.   The larger $l$ values are underestimated (i.e. entanglement is underestimated) particularly for calculations with smaller number of states in the blocks.  Of course, the energy is also less accurately calculated in the second Landau level by dmrg.  This is not the whole story, since the 800 state calculation in the second Landau level  does better for the energy  (on a relative basis) then the 200 state calculation in the first Landau level.  However, the 200 state calculation still does better in calculating $S(l)$.  

Even though it seems possible to use more states in the blocks (reference \cite{feiguin} uses up to 5000) it appears to be difficult to go much  beyond direct diagonalization in calculating the entanglement entropy in the second Landau level.   A simple estimate shows, based on the above calculations,  why this is the case.  The computation  for $\nu=7/3$ indicates at least 1000 block states (and this may be an under estimate) are necessary to get a fairly accurate result.  In going from 36 to 48 sites (12 to 16 electrons) the "worst" block goes from 18 to 24 sites (1/2 the system size, since the entanglement entropy of the system and environment are equal).  Assuming that the number of states kept needs to scale as     $e^{c \sqrt{s}}$, the number of states needed for 48 sites is at least $1000^{\sqrt{24/18}} \approx$ 3000 states.

\section{Conclusion}
The entanglement entropy of the     $\nu=1/3$ and   $\nu= 5/2$ quantum Hall states in the presence of short range disorder has been calculated  by direct diagonalization.  For very weak disorder, the value of the topological entanglement entropy (  a universal  quantity)  is roughly consistent with the expected  theoretical results and disorder free calculations.  However, ( in particular for   $\nu= 5/2$) the advantages of having less system size dependence with weak disorder are outweighed by the disadvantage of the inaccessibility of larger system  sizes.  To investigate the possibility of using the entanglement entropy to detect quantum phase transitions, the entanglement entropy has been calculated for a broader  range of disorder strength.  For   $\nu= 1/3$ , the $l$-orbital entanglement entropy (figures 7a,b) shows a strong decrease, and the variance (figure 10) shows a strong increase through the range $U_R \approx 0.1 - 0.25$. For the range of  disorder considered and the amount of averaging done, we suggest that this is a possible signature of a phase transition similar to that observed for the mobility gap in reference \cite{wan} at $U_R \approx 0.25$.   For $\nu  = 5/2$ we see a much sharper transition feature in the $l$-orbital entanglement entropy (figures 8a,b) and at a much smaller  value of the disorder strength, $U_R \approx 0.04$. Despite the sharper transition, there is no corresponding feature in the variance (figure 9), as there is in the $\nu = 1/3$ case. The sensitivity of the 5/2  state to disorder is well known from experimental studies where samples must have a high (zero  field) mobility to see an incompressible state.    Thus there is  qualitative agreement with experiment, taken with due caution in that a quantitative comparison likely requires considering longer  range disorder.    In our study, one number, the entanglement entropy has been used to characterize the reduced density matrix.  There is possibly additional information in the full spectrum of the reduced density matrix \cite{li}, which has been shown to be related to the conformal field theory describing the one dimensional edge state of the quantum Hall state \cite{li,zozulya2,ivan}.  It would definitely be of interest \cite{refael} to study the entanglement spectrum in the present system.  Even if the topological entanglement entropy (derived from the entanglement entropy) is a complete invariant  \cite{fla}, numerically it may well be easier to see transitions using the entire spectrum \cite{li,zozulya2}.  Finally, we have displayed some preliminary results  using dmrg to compute the entanglement entropy.  These results indicate dmrg holds some promise in calculating the entanglement entropy in the lowest Landau level; it appears more difficult to do calculations in the second Landau level and to go much beyond systems that one can treat by direct diagonalization.  This may indicate that potentially more powerful numerical methods, for example, tensor network states \cite{corboz} or the methods of reference \cite{thomale2}, will prove useful.  

\section{Final Remarks}

After this manuscript was posted at arXiv.org, we became aware of an interesting paper that calculates the topological entanglement entropy using a different method in the flat torus geometry. (We thank Dr. Haque for bringing this reference to our attention.)  In essence, ref. \cite{lauchli} , calculates the entanglement entropy $S(N/2)$ taking the subsystem to be half the system size.  The scaling law $S(N/2) \sim c_1 \sqrt{\frac{N}{\alpha}} - 2\gamma$ is then used where $\alpha$ is the aspect ratio and $N$ is the number of orbitals in the system; this approach was also used by Shibata \cite{shibata3}.  In the method described in section II (see also \cite{ent1,ent2} ) the scaling law $S(l) \sim c_2 \sqrt{l} - 2 \gamma $ is used where $l$  the number of orbitals in the subsystem is much smaller then $N$.  In this method, the subsystem for fixed $l$ becomes increasingly thin since the number of  states per unit length (the magnetic length) scales as $ \sqrt{N} $.   

Let us examine this point \cite{lauchli} in greater detail.  Imagine there is a subsystem consisting of a fixed number of orbitals $l$ and $N$ becomes very large.  Consider the square torus geometry, a "box" of dimensions $a X a$; here $a = \sqrt{2 \pi N}$.  The width of a box of $l$ orbitals is $\frac{l}{N} \sqrt{2 \pi N}$ i.e. $l \sqrt{\frac{2\pi}{N}}$ so the width goes to zero as $\sqrt{\frac{1}{N}}$.  However, at the same time the width goes to zero, the length goes as $\sqrt{2\pi N}$.  Although the width and length are both "singular" as $N$  goes to infinity, the area is perfectly well defined, $2\pi l$ (again in units of the magnetic length squared).  Since the area law relates the entanglement entropy to a linear dimension of the subsystem, it is reasonable that $S(l)$ scales as the square root of the area,  $  S(l) \sim c\sqrt{l}$, and this is verified by explicit calculations.

It should be emphasized that neither approach is fully justified by the considerations in ref. \cite{kitaevpreskill,levinwen}.  At least for the current state of knowledge, the best justification for either method is that they give reasonable results where the physics is well understood, Laughlin states.  This is true for both techniques, hence in principle, either technique can be used to calculate the topological entanglement entropy.  That being said, since system sizes are limited, one technique may well be superior depending on the filling fraction in question.
In particular, the method of reference \cite{lauchli,shibata3} allows one to extract $\gamma$ more accurately for $\nu=1/3$ from finite size calculations.  As an illustration of this method, in figure 13, $S(17)$ is plotted vs. $\sqrt{\frac{N}{\alpha}}$ for 11 electrons in 33 states in the lowest Landau level (no disorder).  For this system size the orbitals begin to strongly overlap at $\sqrt{\frac{N}{\alpha}} = \sqrt{2\pi} \approx 2.51$.  (At this value of $\sqrt{\frac{N}{\alpha}}$ the orbitals are one magnetic length apart).  If one uses this value as a cutoff and fits the $S(17)$ curve to a line for  $\sqrt{\frac{N}{\alpha}} > \sqrt{2\pi}$ a topological entanglement entropy $\gamma$ of $1.14 \pm .02$ is obtained.  As a comparison $S(17)$ vs. $\sqrt{\frac{N}{\alpha}}$ is plotted in figure 14 for the same filling in the second Landau level.  This plot is calculated from the lowest energy states in the momentum sectors $k_y =11,22,33$ which are the ground states at aspect ratio one.  There is evidence of several transitions as the thin torus is transformed to a square.  It is interesting that the transitions in this figure as a function of   $\sqrt{\frac{N}{\alpha}}$ bear some resemblance to the "plateau" transition as a function of disorder strength as seen in figure 8.    

This work was supported  in part by NSF Grant no. 0705048 (B.F. and D. L.) and the Department of Energy, DE-FG02-08ER64623---Hofstra University Center for Condensed Matter (G. L.).

\bigskip
\bigskip

\noindent
{\bf Figure Captions}
\bigskip

\noindent
figure 1.  Entanglement entropy vs. $\sqrt{l}$ for 10 electrons in 30 orbitals.  The green circles are for no disorder, while the blue and red circles are for $U_R$ =.05 averaged over 10 and 100 samples respectively.  The error bars are given by the root mean square values of $S(l)$.
\bigskip

\noindent
figure 2.  Linear fit to the initial part of the $<S(l)>$ vs. $\sqrt{l}$ curve. $U_R$=.05 and the filling $\nu=1/3$ .
\bigskip

\noindent
figure 3. Dependence of topological entanglement entanglement entropy $\gamma$ on system size for filling 1/3.  $U_R$=.01 and $\gamma$ is plotted vs. $1/N$.  ($N$=number of orbitals).
\bigskip

\noindent
figure 4. Extrapolated entanglement entropy $<S^*(l)>$ vs. $\sqrt{l}$ for $\nu=1/3$.  Systems with 21 to 30 orbitals were used in the extrapolations and $U_R$=.01. 
\bigskip

\noindent
figure 5. Dependence of the topological entanglement entropy $\gamma$ on system size for filling factor $5/2$.  $U_R$=.01 and $\gamma$ was obtained from the initial slope of $<S(l)>$.
\bigskip

\noindent
figure 6. Extrapolated entanglement entropy $<S^*(l)>$ vs. $\sqrt{l}$ for $\nu=5/2$.  Systems of sizes 24,26, 28 were used in the extrapolations and $U_R$ = .01.
\bigskip

\noindent
figure 7. $<S(l)>$ vs. $U_R$ for filling $1/3$.  In figures 7a  $<S(2)>$ is  plotted, while figure 7b is a plot of  $<S(12)>$.
\bigskip

\noindent
figure 8. $<S(l)>$  vs. $U_R$ for filling $5/2$.  In figures 8a $<S(2)>$ is  plotted, while figure 8b is a  plot of  $<S(12)>$.
\bigskip

\noindent
figure 9. The variance of $S(12)$, $<S(12)^2>-<S(12)>^2$ vs. $U_R$ for filling $5/2$. 
\bigskip

\noindent
figure 10. The variance of $S(12)$, $<S(12)^2>-<S(12)>^2$ vs. $U_R$ for filling $1/3$.  
\bigskip

\noindent
figure 11. $S(l)$ vs. $\sqrt{l}$ for $\nu=1/3$, calculated by the density matrix renormalization group (dmrg).  The different symbols correspond to different number of states in the dmrg blocks.  The green squares are the exact results.
\bigskip

\noindent
figure 12. $S(l)$ vs. $\sqrt{l}$ for $\nu=7/3$, calculated by dmrg.  The different symbols correspond to different number of states in the dmrg blocks.  The green squares are the exact results, while the green crosses are values extrapolated from dmrg.
\bigskip

\noindent
figure 13. $S(17)$ vs.  $\sqrt{\frac{N}{\alpha}}$ for $\nu=1/3$, 11 electrons in 33 orbitals.  The line is a fit to $S(17)$ for $\sqrt{\frac{N}{\alpha}} > \sqrt{2\pi}$.
\bigskip

\noindent
figure 14.  $S(17)$ vs.  $\sqrt{\frac{N}{\alpha}}$ for $\nu=7/3$, 11 electrons in 33 orbitals.

\eject

\begin{figure}[ht]
\includegraphics[width=14cm]{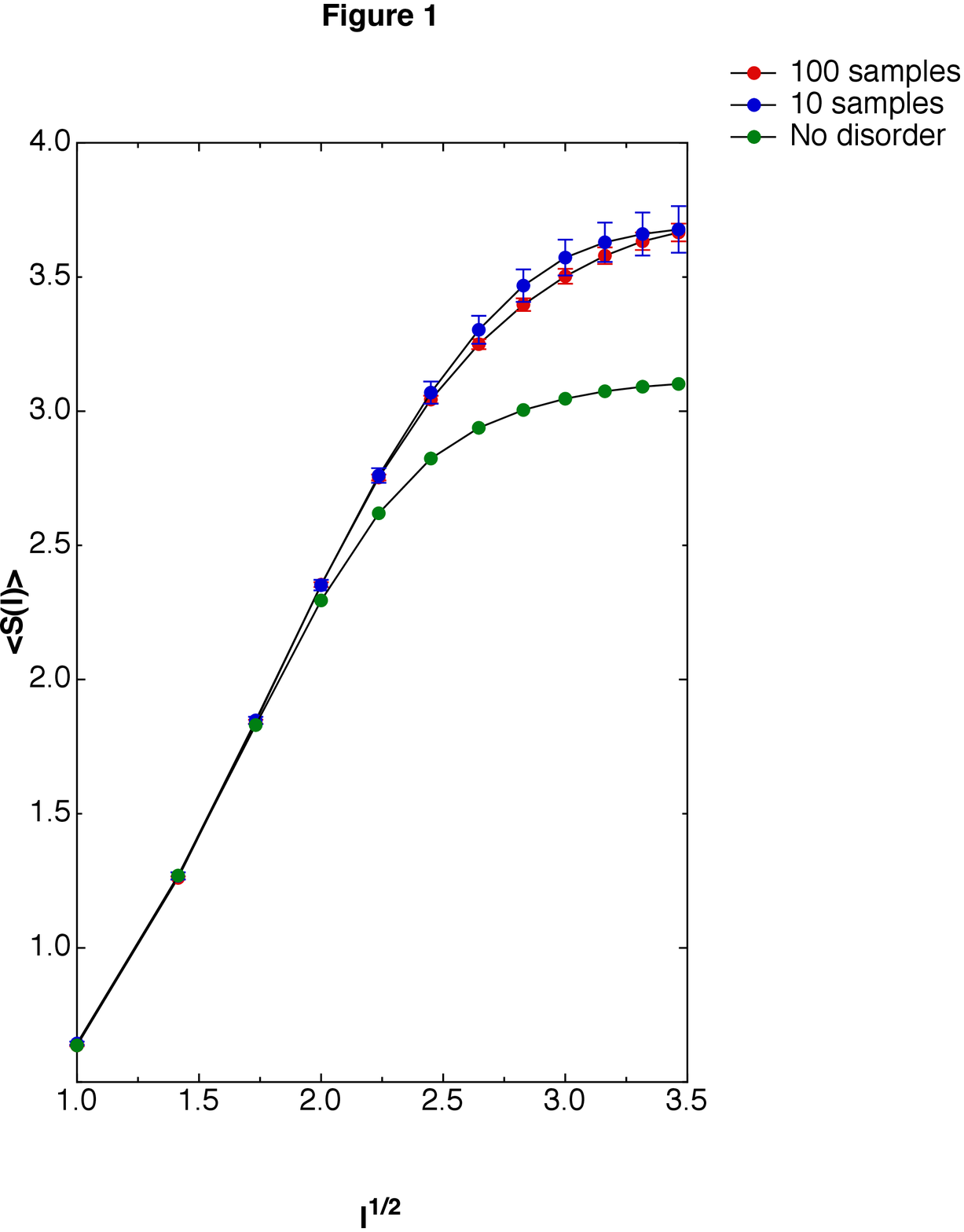}
\end{figure}

\begin{figure}[ht]
\includegraphics[width=14cm]{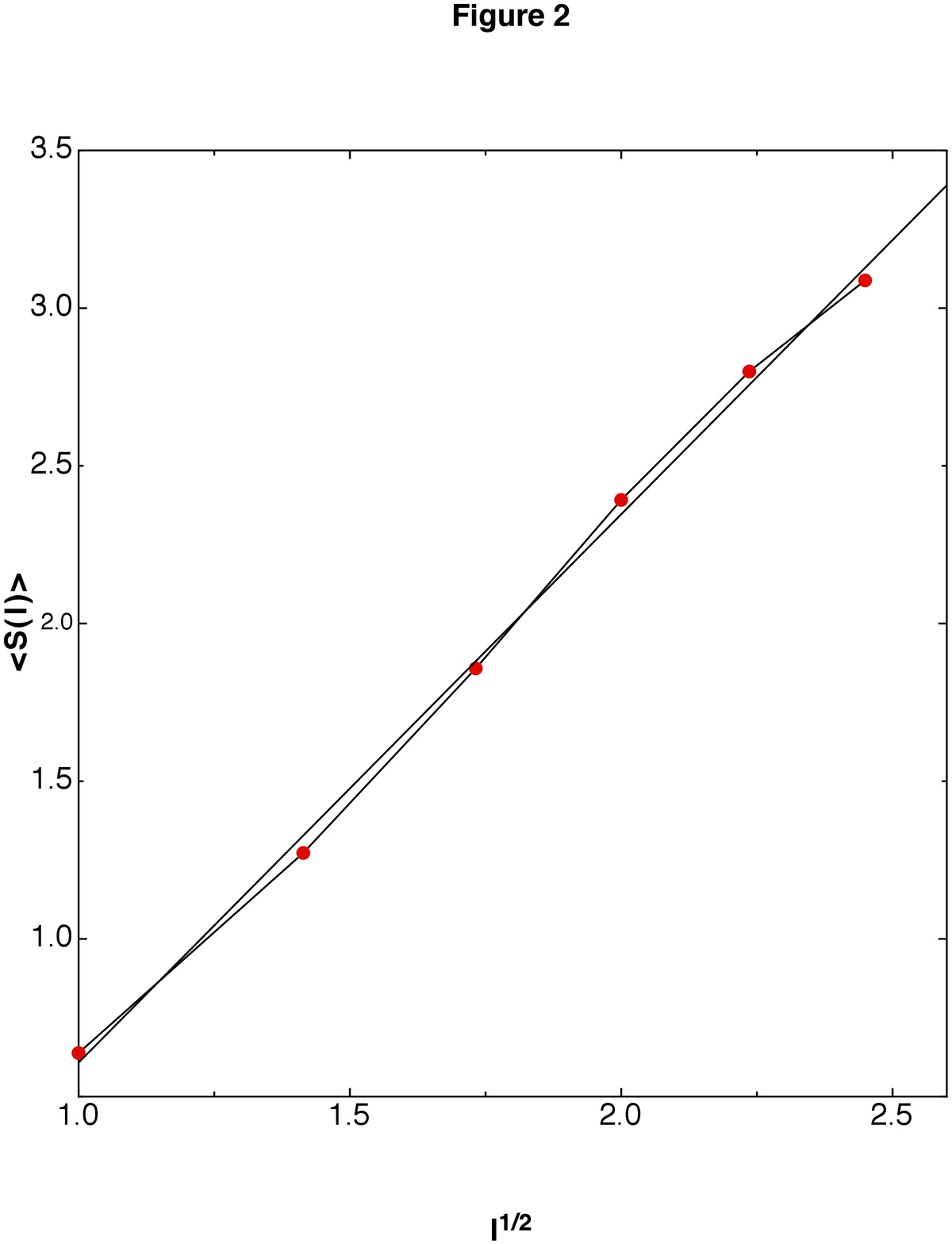}
\end{figure}

\clearpage

\begin{figure}[ht]
\includegraphics[width=14cm]{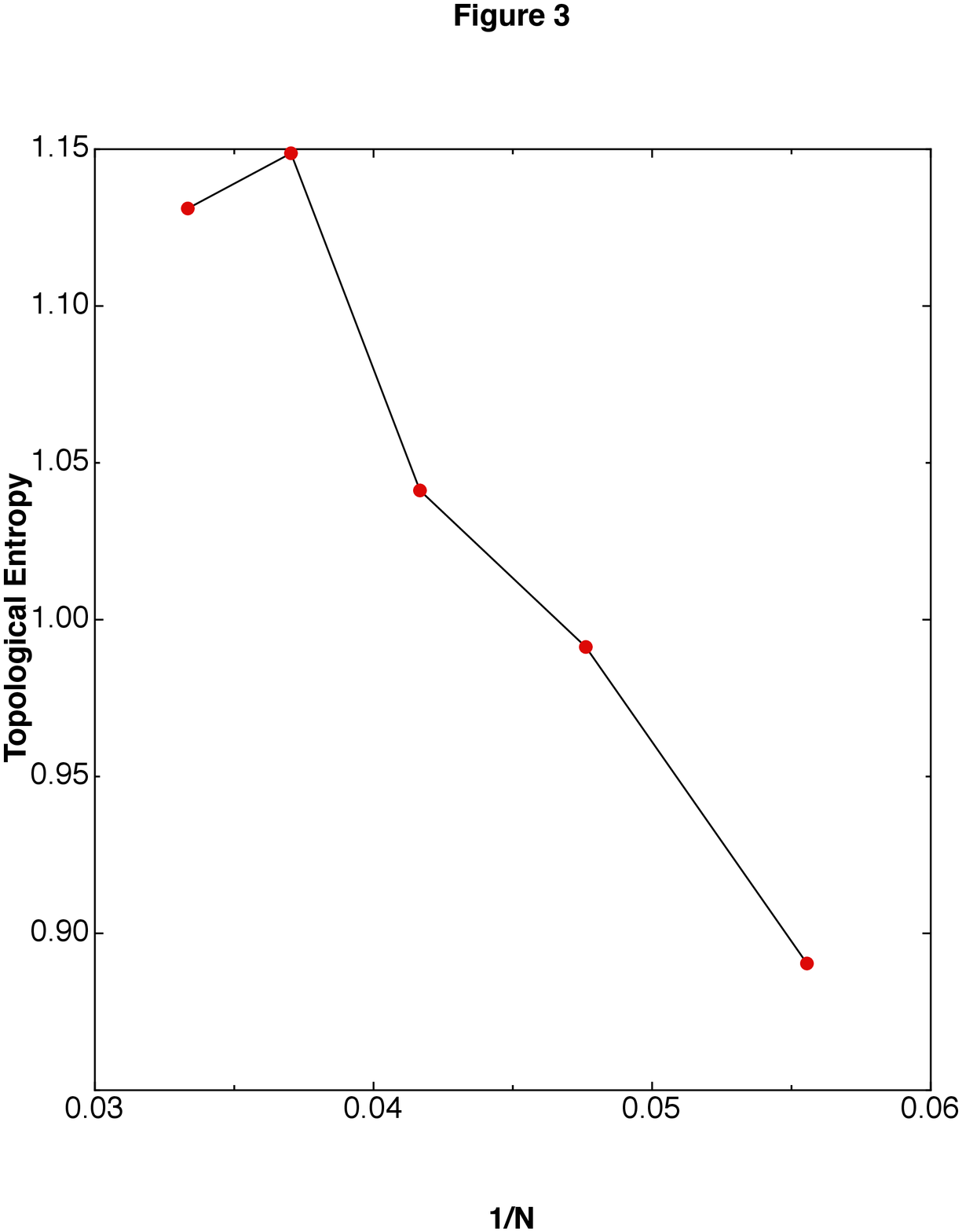}
\end{figure}

\begin{figure}[ht]
\includegraphics[width=14cm]{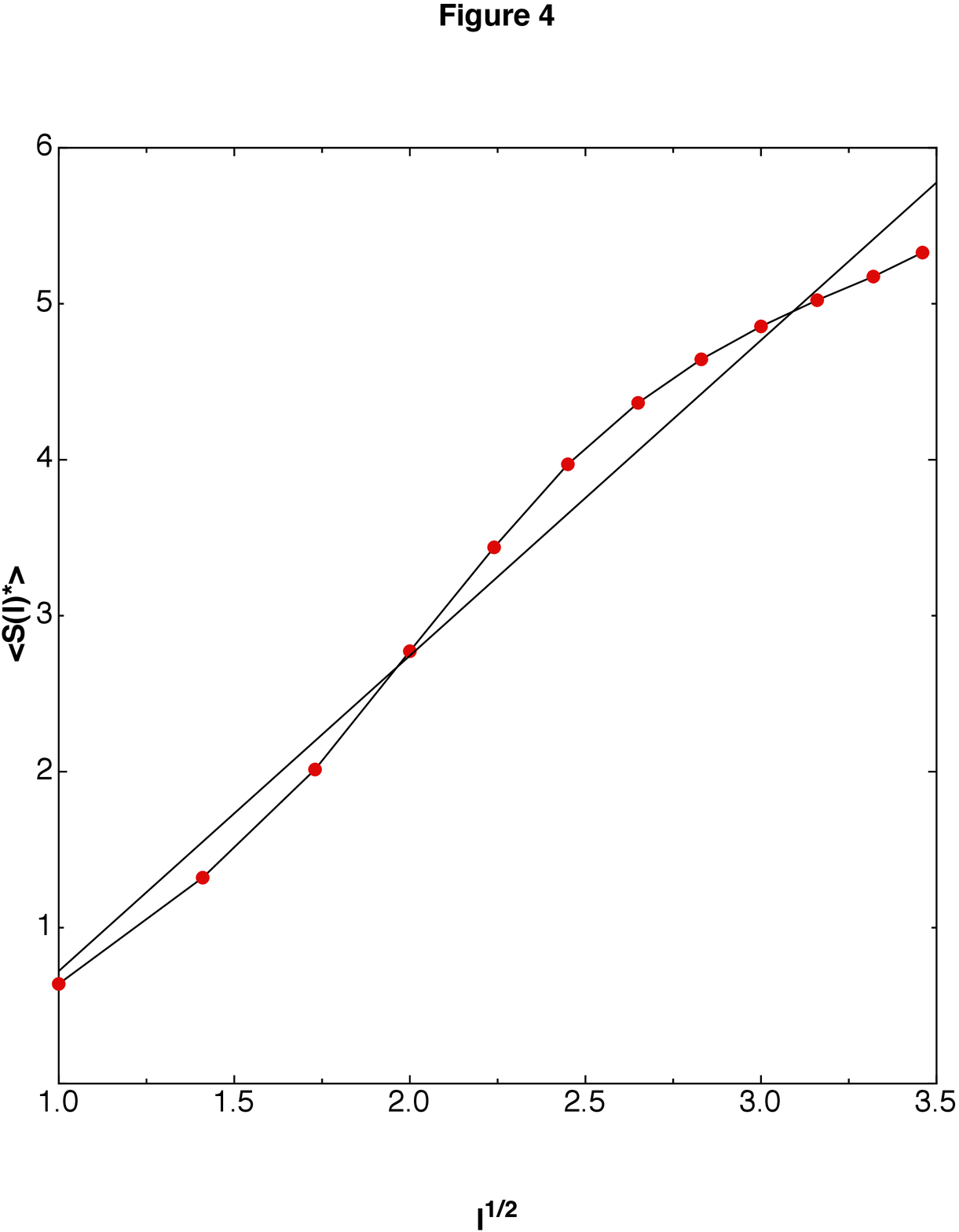}
\end{figure}

\begin{figure}[ht]
\includegraphics[width=14cm]{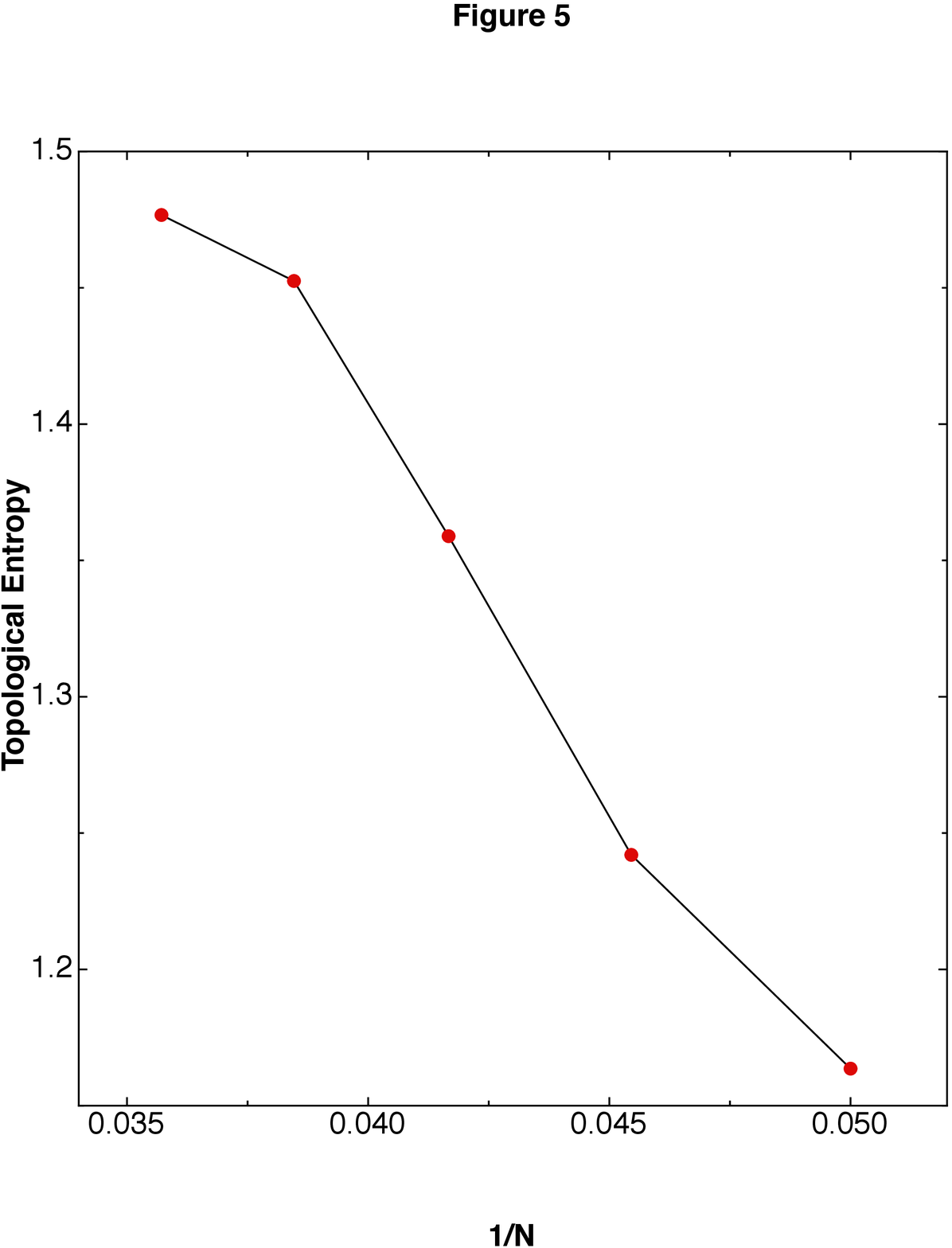}
\end{figure}

\begin{figure}[ht]
\includegraphics[width=14cm]{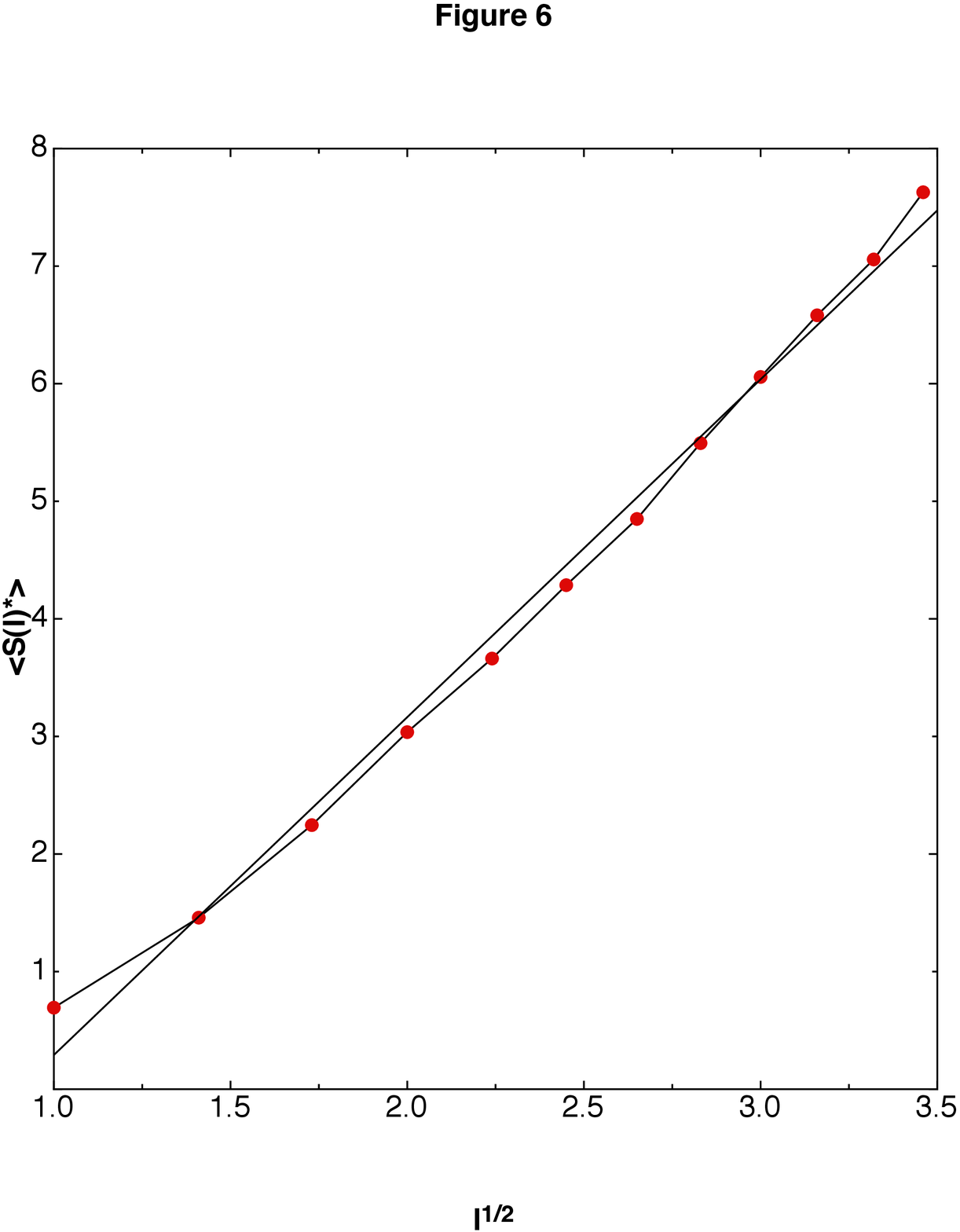}
\end{figure}

\begin{figure}[ht]
\includegraphics[width=14cm]{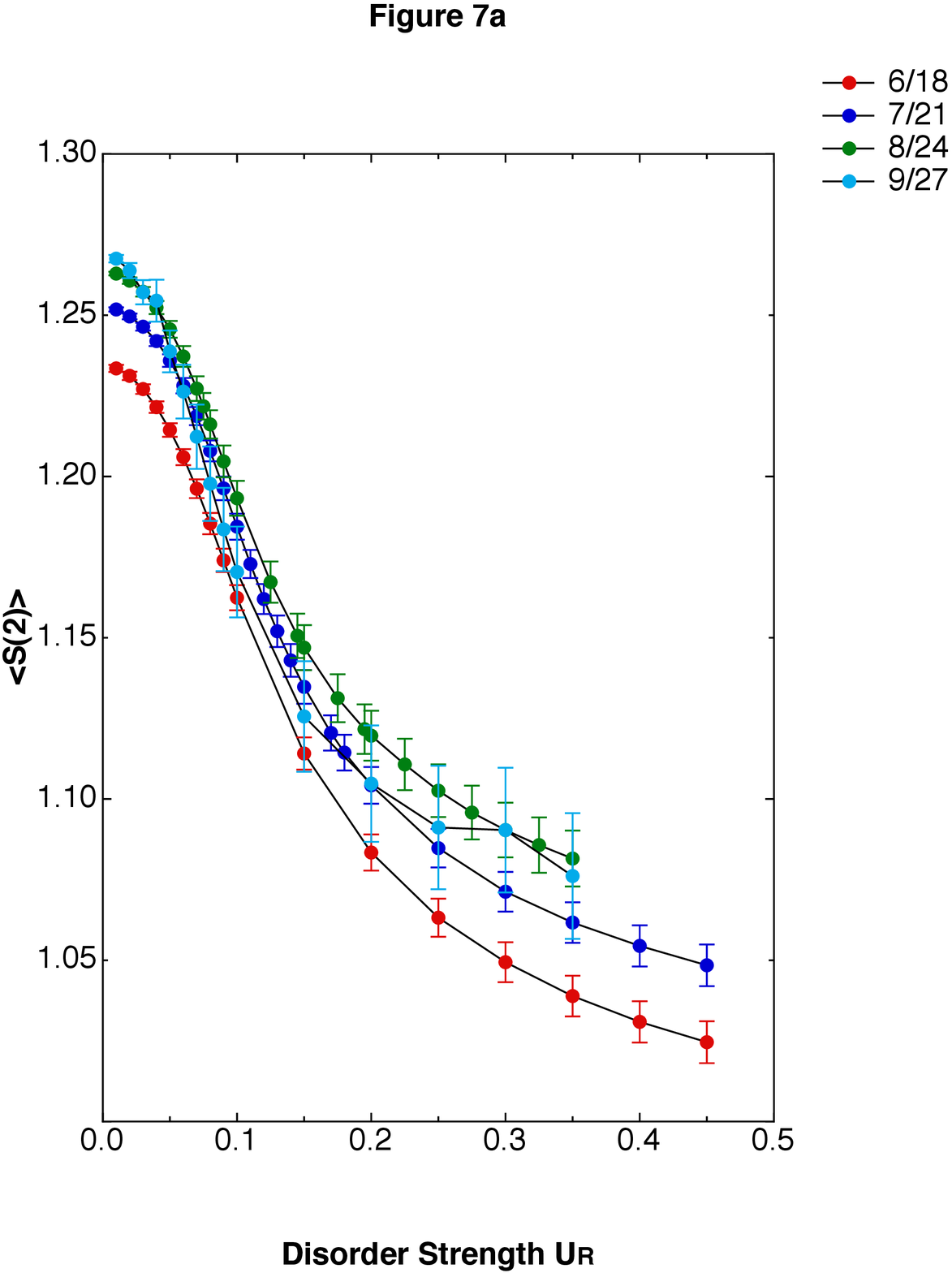}
\end{figure}

\begin{figure}[ht]
\includegraphics[width=14cm]{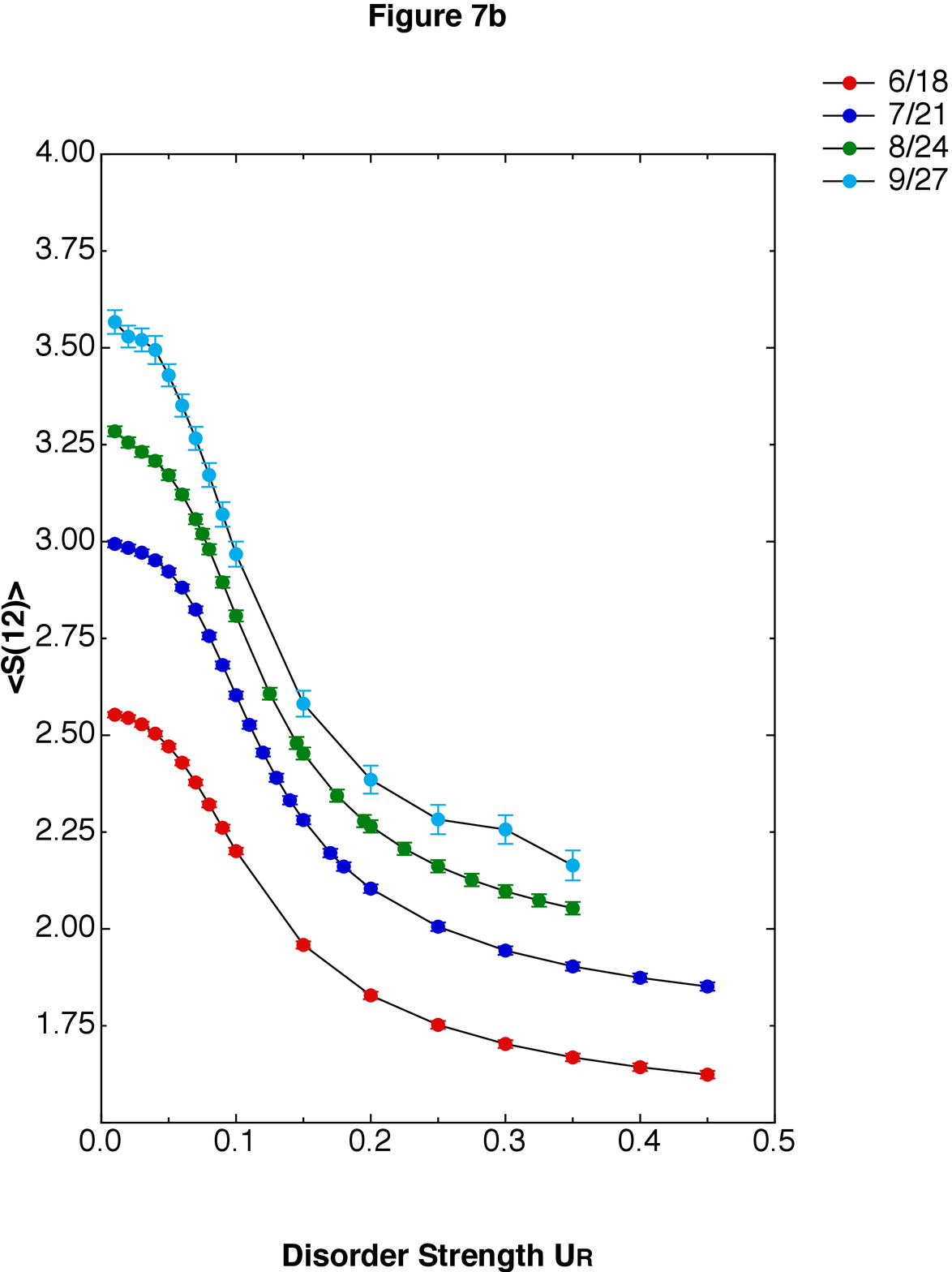}
\end{figure}

\begin{figure}[ht]
\includegraphics[width=14cm]{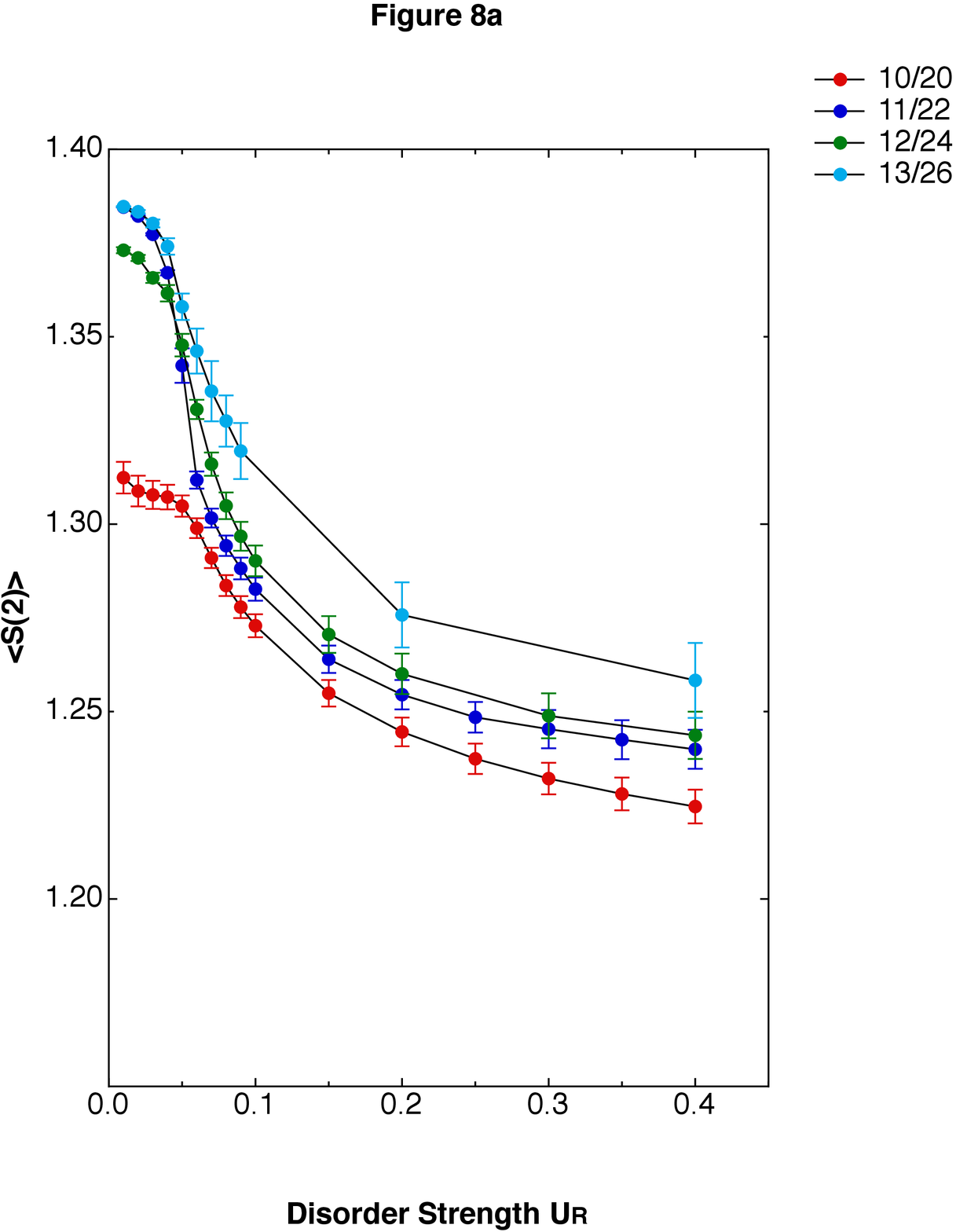}
\end{figure}

\begin{figure}[ht]
\includegraphics[width=14cm]{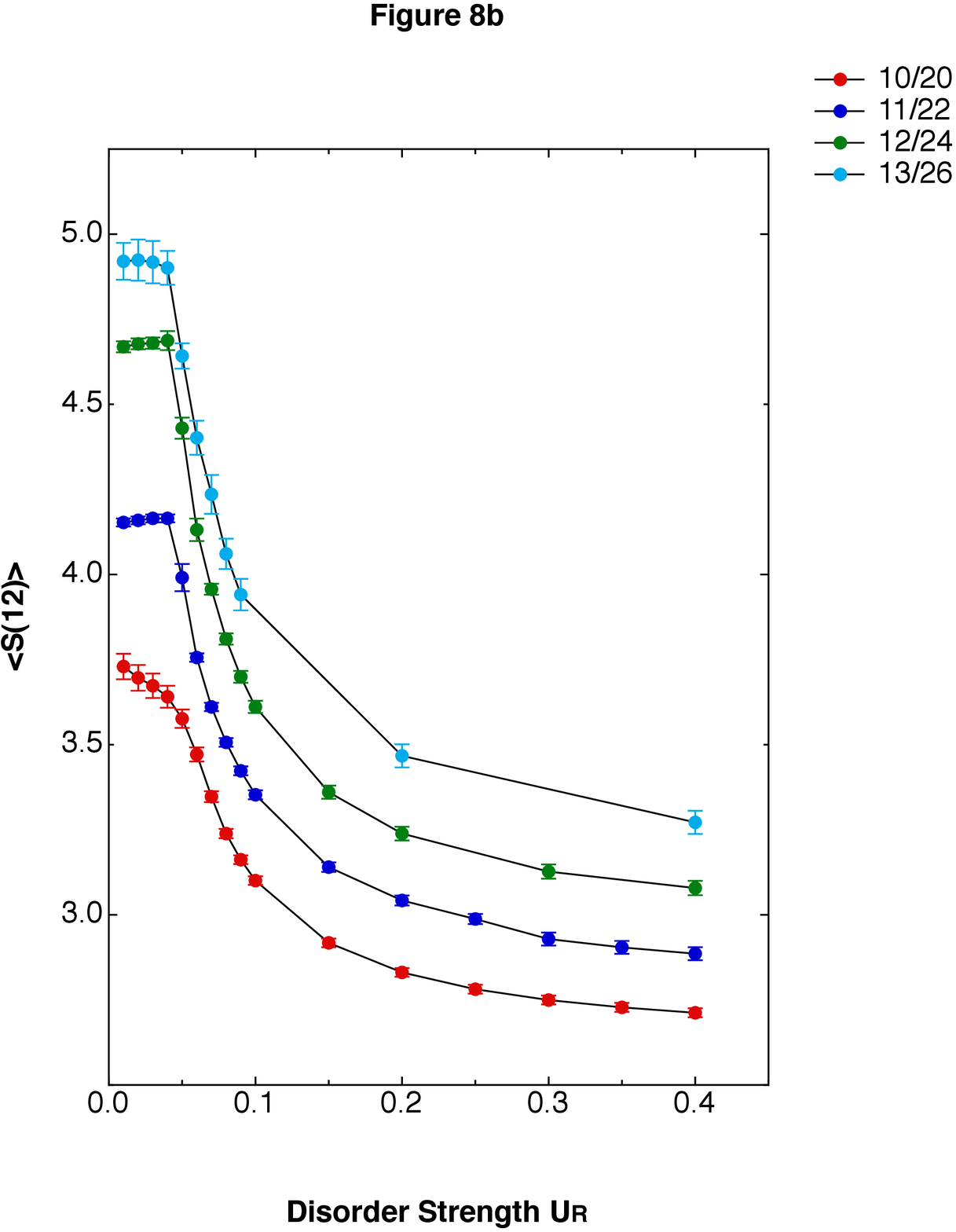}
\end{figure}

\begin{figure}[ht]
\includegraphics[width=14cm]{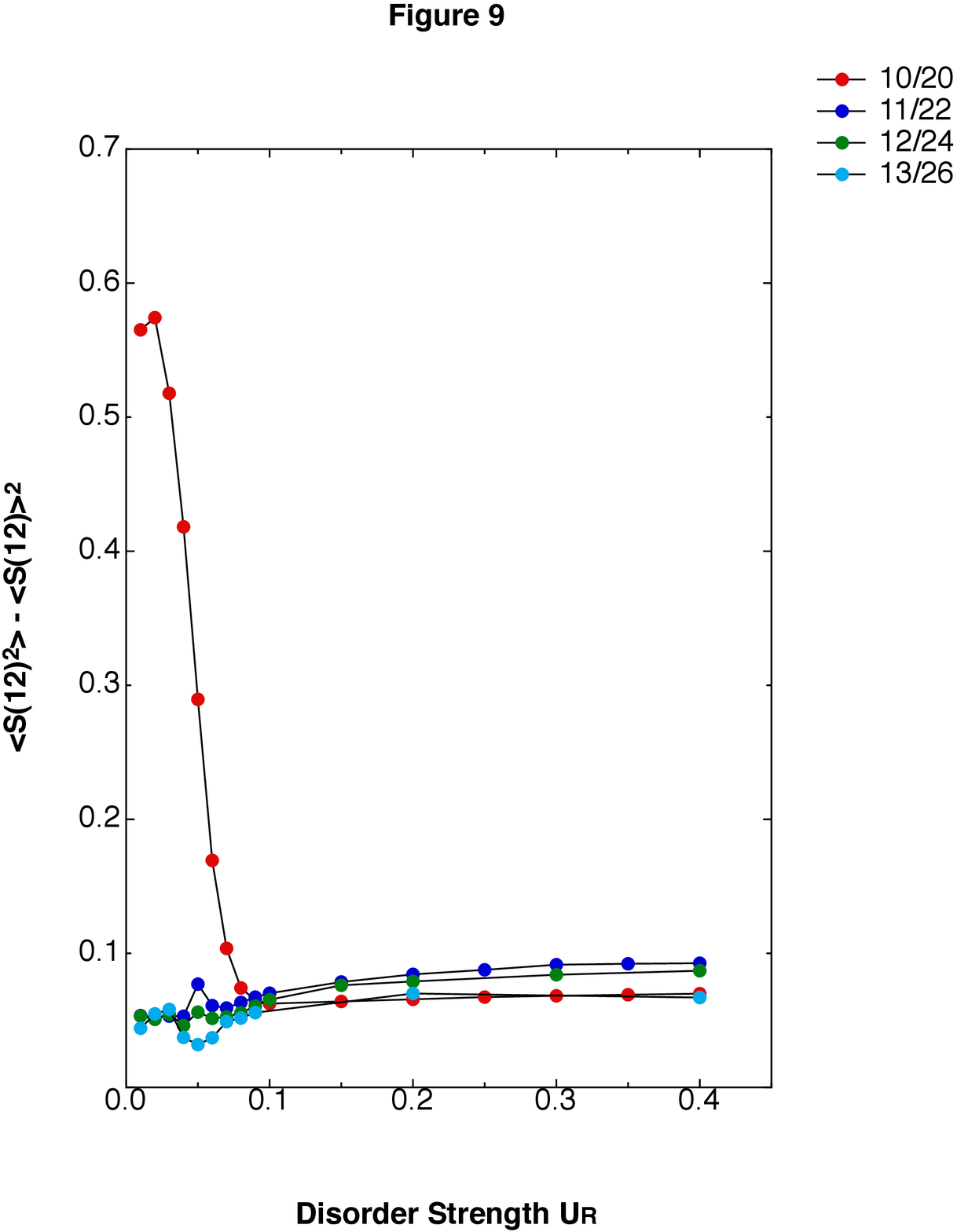}
\end{figure}

\begin{figure}[ht]
\includegraphics[width=14cm]{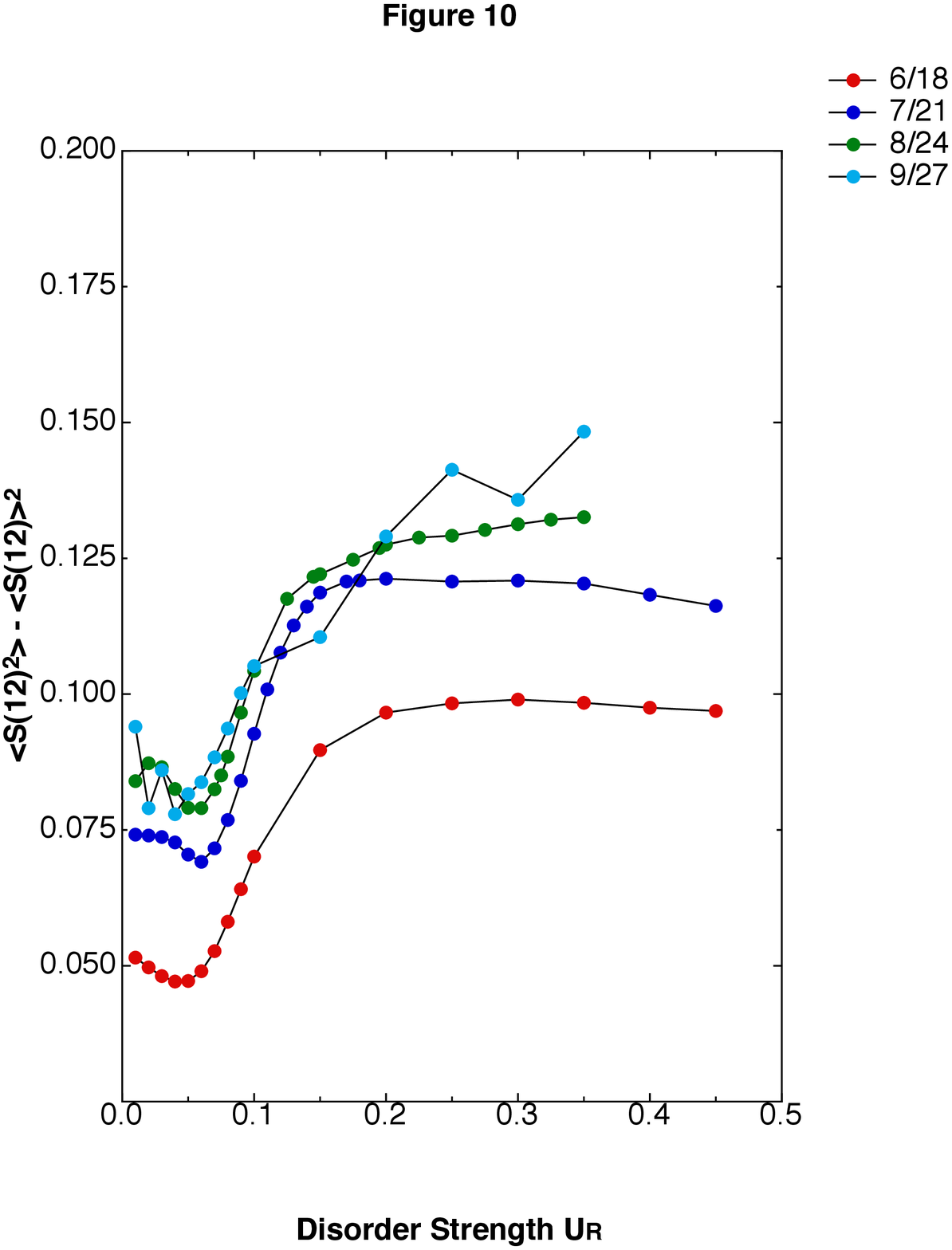}
\end{figure}

\begin{figure}[ht]
\includegraphics[width=14cm]{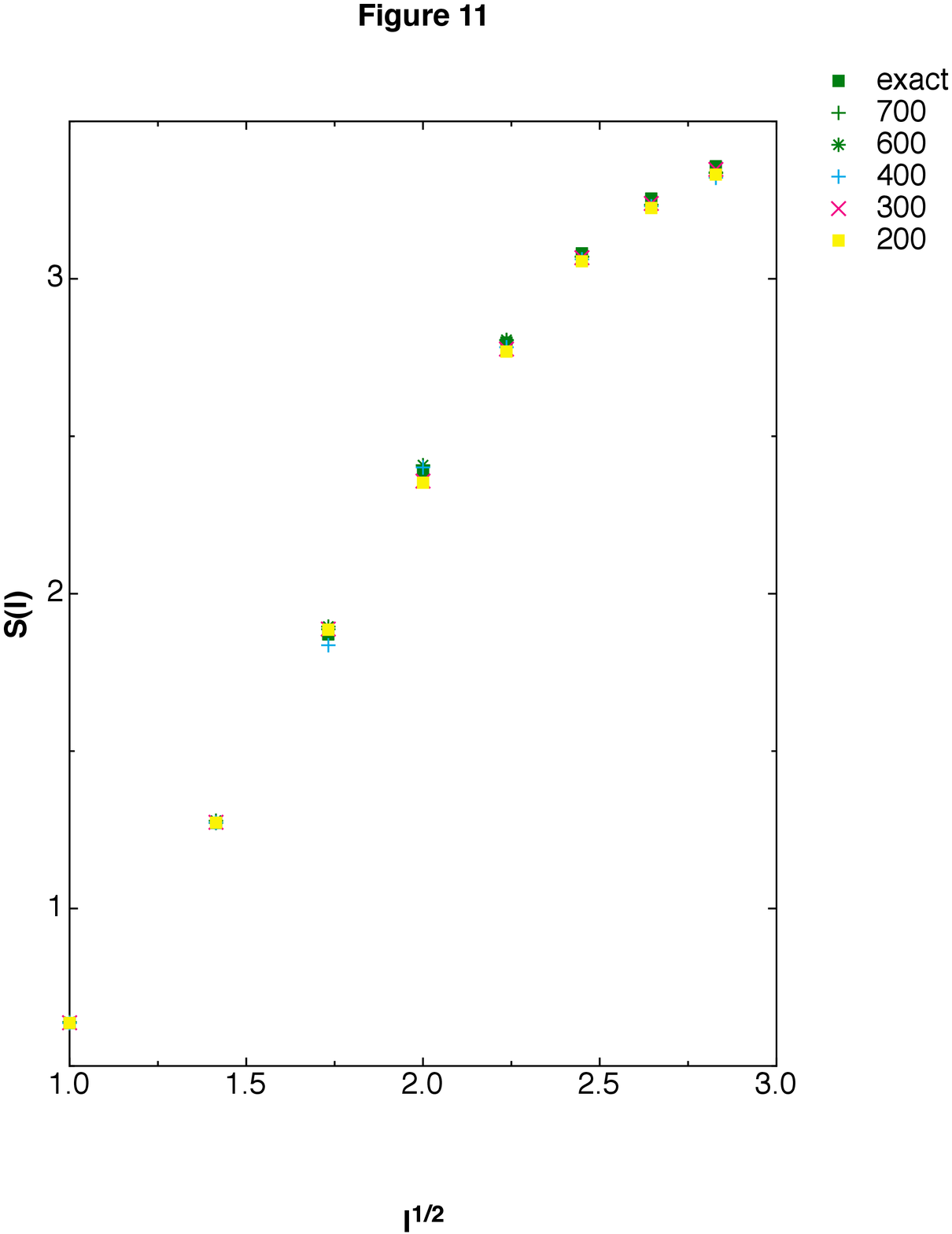}
\end{figure}

\begin{figure}[ht]
\includegraphics[width=14cm]{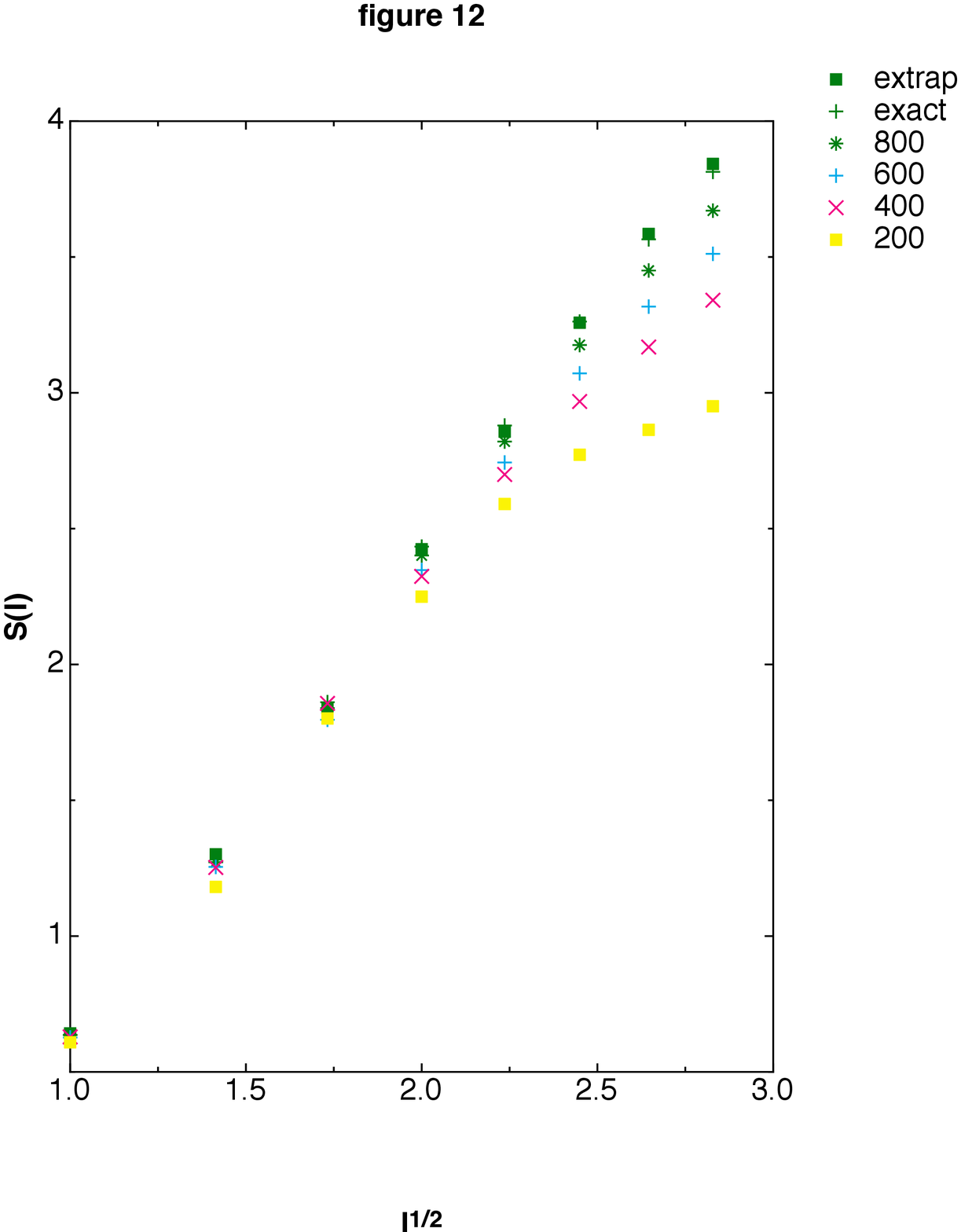}
\end{figure}

\begin{figure}[ht]
\includegraphics[width=14cm]{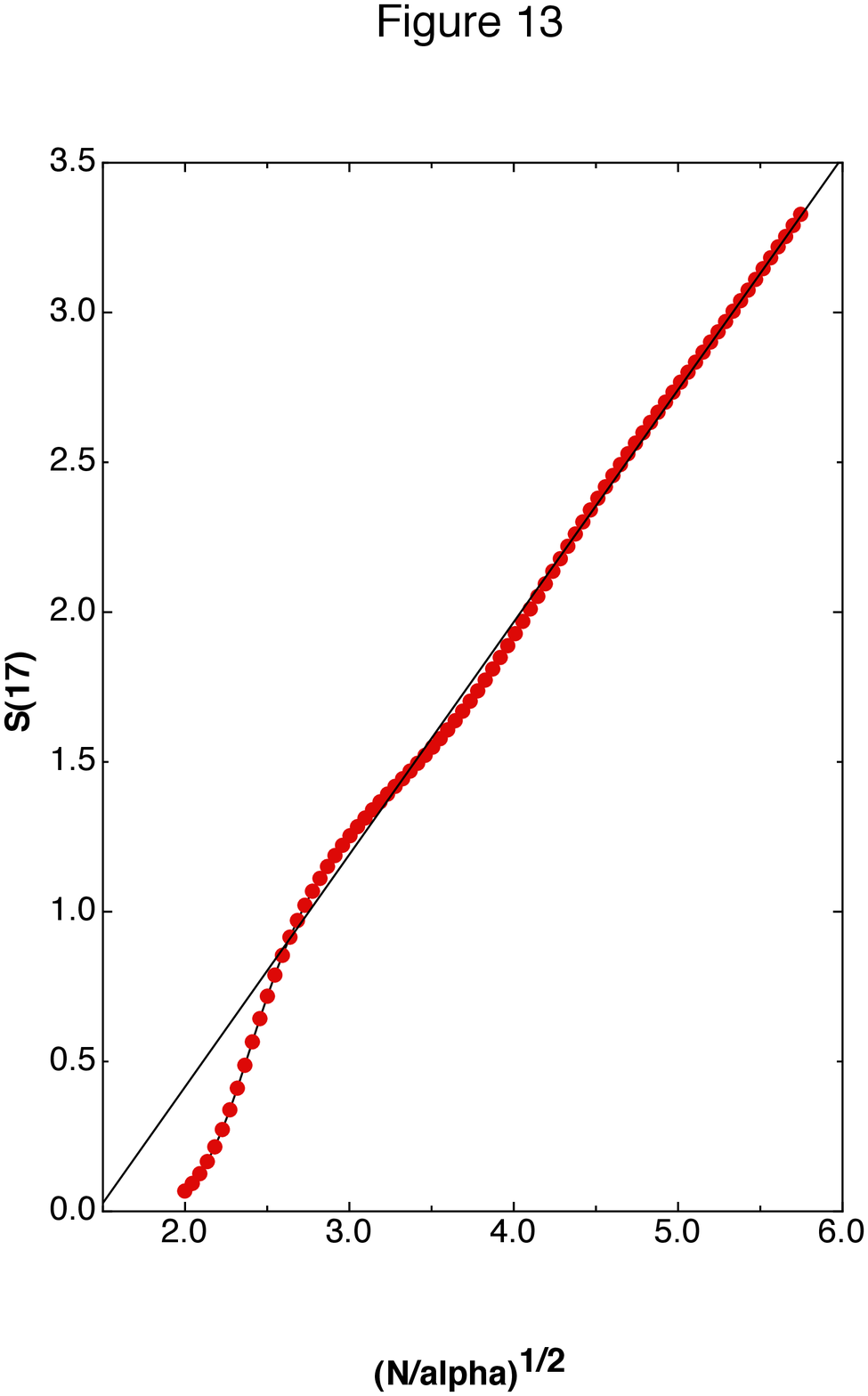}
\end{figure}

\begin{figure}[ht]
\includegraphics[width=14cm]{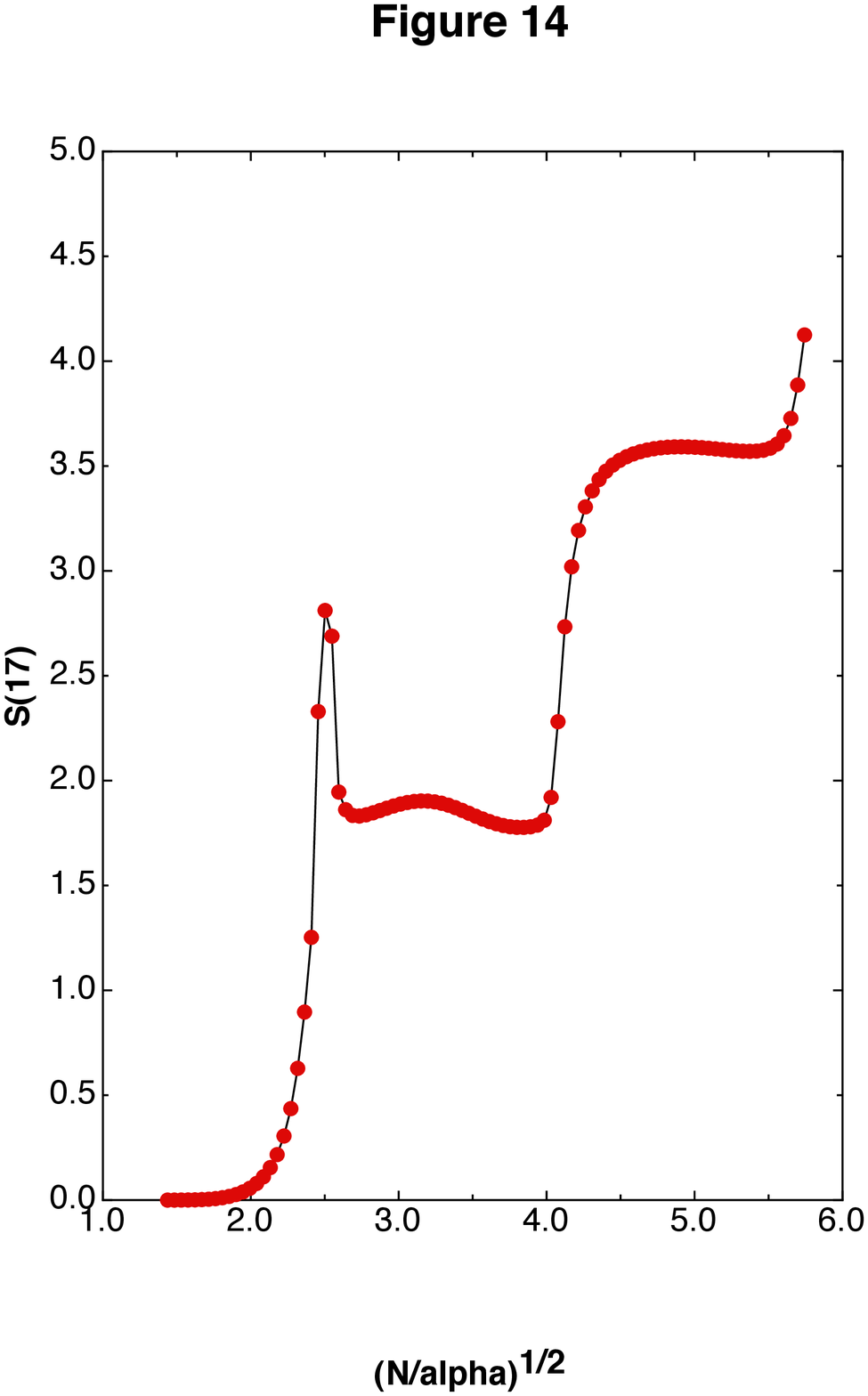}
\end{figure}


\begin{references}

\bibitem{kitaevpreskill} A. Kitaev and J. Preskill, Phys. Rev. Lett. 96, 110404 (2006).

\bibitem{levinwen}  M. Levin and X.-G. Wen, Phys. Rev. Lett. 96, 110405 (2006).

\bibitem{haque} M. Haque, O. Zozulya and K. Schoutens, Phys. Rev. Lett. 98, 060401 (2007).

\bibitem{zozulya} O. S. Zozulya, M. Haque, K. Shoutens, and E. H. Rezayi, Phys. Rev. B {\bf 76}, 125310 (2007).

\bibitem{ent1} B. A. Friedman and G. C. Levine, Phys. Rev. B {\bf 78}, 035320 (2008).

\bibitem{morris} A. G. Morris and D. L. Feder, Phys. Rev. A {\bf 79}, 013619 (2009).

\bibitem{ent2} B. A. Friedman and G. C. Levine, Int. J. Mod. Phys. B to be published,arXiv: 0902.1524.

\bibitem{zozulya2} O. Zozulya, M. Haque, and N. Regnault, Phys. Rev. B 79, 045409 (2009).

\bibitem{nayak} C. Nayak, S. Simon, A. Stern, M. Freedman, and S. Das Sarma, Rev. Mod. Phys. 80, 1083 (2008).

\bibitem{ultrav} J. Dumoit and B. Friedman, J. Phys.: Condens. Matter  16, 3663 (2004); B. Friedman and B. McCarty, J. Phys. : Condens. Matter 17, 7335 (2005).

\bibitem{refael} G. Refael and J. E. Moore, arXiv:0908.1986v1.

\bibitem{chak} S. Chakravarty, arXiv:1004.0730v1.

\bibitem{mund} C. Mund, O. Legeza and R. M. Noack arXiv: 0904.4673.

\bibitem{li} H. Li and F. D. M. Haldane, Phys. Rev. Lett. 101, 010504 (2008).

\bibitem{white} S. R. White, Phys. Rev. B, 10354 (1993).

\bibitem{shibata1} N. Shibata and D. Yoshioka, Phys. Rev. Lett. 86, 5755 (2001).

\bibitem{shibata2} N. Shibata, J. Phys. A 36 R381 (2003).

\bibitem{feiguin} A. E. Feiguin, E. Rezayi, C. Nayak and S. Das Sarma, Phys. Rev. Lett. 100, 166803 (2008).

\bibitem{withrow} B. Friedman and C. Withrow Physica B 403 1500 (2008).

\bibitem{shibata3} N. Shibata, private communcation and March meeting APS Portland, 2010.

\bibitem{corboz} P. Corboz et al. arXiv:0912.0646.

\bibitem{wan} Xin Wan, D. N. Sheng, E. H. Rezayi, Kun Yang, R. N. Bhatt, and F. D. M. Haldane, Phys. Rev. B 72, 075325 (2005).

\bibitem{lauchli} A. M. Lauchli, E. J. Bergholtz, and M. Haque, New J. Phys. 12 (2010) 075004 . 

\bibitem{ivan} I. D. Rodriguez and G. Sierra, arXiv: 1007.5356v1.

\bibitem{fla} S. T. Flammia, A. Hamma, T. L. Hughes, and X.-G. Wen, Phys. Rev. Lett. 103, 261601 (2009).

\bibitem{fend} P. Fendley, M. P. A. Fisher and C. Nayak, J. Stat. Phys. 126, 1111 (2007).

\bibitem{thomale1} R. Thomale, A. Sterdyniak, N. Regnault and B.A. Bernevig, Phys. Rev. Lett. 104, 180502 (2010).

\bibitem{thomale2} R. Thomale, B. Estienne, N. Regnault, and B. A. Bernevig, arXiv: 1010.4837.

\bibitem{read} N. Read and D. Green, Phys. Rev. B 61(15), 10267 (2000).

\bibitem{mac} A. P. Mackenzie et al., Phys. Rev. Lett. 80, 161 (1998).

\end{references}
\end{document}